\documentclass[twocolumn,showpacs,preprintnumbers,amsmath,amssymb]{revtex4}

\usepackage{graphicx}
\usepackage{dcolumn}
\usepackage{bm}

\begin{document}

\preprint{APS/123-QED}

\title{Vertical Discontinuities in Self-Affine Surfaces Lead to Multi-affinity}

\author{S.~J.\ Mitchell}
 \email{S.Mitchell@tue.nl}
 \affiliation{
Schuit Institute for Catalysis
and Department of Chemical Engineering,
Eindhoven University of Technology,
5600 MB Eindhoven, The Netherlands
}

\date{\today}

\begin{abstract}
Many systems of both theoretical and applied interest display multi-affine scaling at small length scales.
We demonstrate analytically and numerically that when vertical discontinuities are introduced
into a self-affine surface, the surface becomes multi-affine.
The discontinuities may correspond to surface overhangs or to an underlying stepped surface.
Two surfaces are numerically examined with different spatial distributions of vertical discontinuities.
The multi-affinity is shown to arise simply from the surface of vertical discontinuities,
and the analytic scaling form at small length scales for the surface of discontinuities is derived
and compared to numerical results.
\end{abstract}

\maketitle

Many systems of both theoretical and applied interest display multi-affine scaling at small length scales
\cite{ALB1,ALB2,ALB3,SDS,CVD,KK}.
Recently, an extensive scaling analysis of surfactant templated hydrogel surfaces 
as measured by atomic force microscopy (AFM) was performed \cite{MK}.
This analysis indicated that the hydrogel surfaces were self-affine;
however, a later numerical study of a frustrated spring-network model 
of cross-linked hydrogels \cite{GB} indicated multi-affine scaling.
Reconciliation of these two observed behaviors led to an interesting and universal conclusion:
introduction of vertical discontinuities into a self-affine surface leads to multi-affine scaling.
To our knowledge, this has not previously been reported in the literature,
most likely because height-height correlations are usually calculated 
for the second power of the height increments,
in which case the surface constructed only of discontinuities resembles a random walk on all length scales.
Here, we provide a discussion which explains this source of multi-affine behavior,
and we present both numerical and analytic results.

Consider a one-dimensional, real, single-valued surface, $z(x)$, 
where $x$ is a real number on the interval, $x \in [0,1]$.
The generalized height-height correlation function \cite{ALB1,ALB2,ALB3,SDS,CVD,KK,GB} for this surface is
\begin{equation}
C_q(r)=\langle |z(x+r)-z(x)|^q \rangle \;,
\label{eq:corr0}
\end{equation}
where $\langle \cdots \rangle$ denotes an average over all $x$ values,
$|r|<1/2$, and $q$ is a positive, non-zero real number.
Without loss of generality, we may assume that $r$ is positive,
since $C_q(r)=C_q(-r)$ for any function $z(x)$.

Often, $C_q(r)$ will display power-law behavior for $r \ll r_{\times}$
and will display a constant value for $r \gg r_{\times}$,
where $r_{\times}$ is some cross-over length scale between the two behaviors.
Surfaces displaying power-law correlations, $C_q(r) = A_q r^{q \alpha_q}$,
fall into one of two categories:
$q$-independent scaling, $\alpha_q=\alpha$, called self-affine scaling,
and $q$-dependent scaling, called multi-affine scaling \cite{ALB1,ALB2,ALB3,SDS,CVD,KK}.

By introducing vertical discontinuities into a self-affine surface,
we can cause the surface to become multi-affine.
Consider the function $z(x)$,
which is self-affine for all $r \ll r_{\times}$.
For the numerical results, self-affine surfaces were generated using the method of Ref.~\cite{VOSS}.
We introduce a finite number, $N$, of vertical discontinuities into the function $z(x)$
such that the new surface is
\begin{equation}
z'(x)=\left\{\sum_{i=1}^N \delta_i\Theta(x-x_i) \right\} +z(x)\; ,
\label{eq:zprime}
\end{equation}
where $i$ indexes the discontinuities,
$\delta_i$ is the magnitude of the discontinuity at $x=x_i$,
and $\Theta(y)$ is a step function which is zero for $y < 0$ and 1 for $y \ge 0$.
Without loss of generality,
we can assume an order to the set $\{x_i\}$ such that $x_{i-1}<x_i$ and $x_0=0$.

\begin{figure}
\includegraphics[width=0.75\columnwidth]{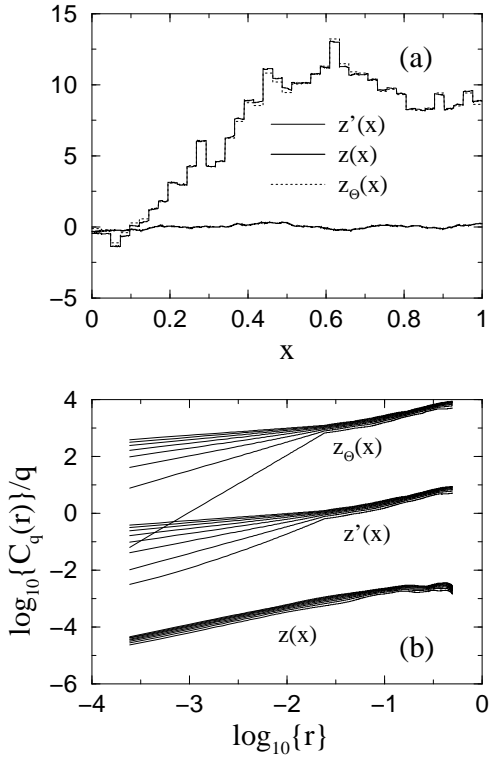}
\caption{Multi-affine surface generated from Eq.~(\ref{eq:zprime}) with discontinuities evenly spaced in $x$.
The magnitudes of the discontinuities, $\delta_i$, are drawn independently from a Gaussian distribution
with standard deviation 1.0 and mean 0,
and $N=41$.
The self-affine function $z(x)$ has $\alpha=0.75$.
(a) The surface, $z'(x)$, and related functions.
(b) $C_q(r)$ for each function shown in (a) as labeled in the plot.
The correlation functions for $z_\Theta(x)$ and $z(x)$ have been displaced up by 3 and down by 2 units, respectively,
for graphical clarity.
No graphical distinction is made between curves with different $q$,
but $q \in [0.5,4.0]$ in steps of 0.5,
and for the curves shown here, $C_q(r)>C_p(r)$ when $q>p$.
The cross-over length scale is $r_{\times}=1/N$,
and the cross-over region is very narrow.
}
\label{fig:even}
\end{figure}

\begin{figure}
\includegraphics[width=0.75\columnwidth]{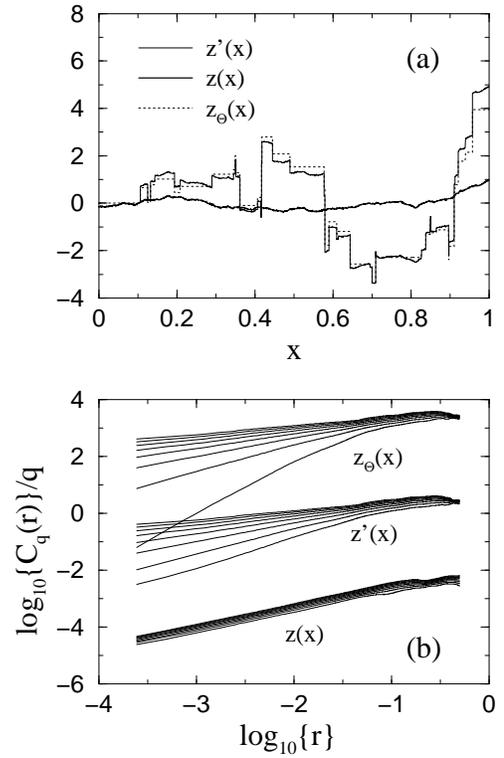}
\caption{Multi-affine surface generated from Eq.~(\ref{eq:zprime}) with discontinuities randomly spaced in $x$.
All parameters are the same as in Fig.~\ref{fig:even}.
(a) The surface, $z'(x)$, and related functions.
(b) $C_q(r)$ for each function shown in (a) as labeled in the plot.
The cross over length scale is $r_{\times}=1/N$,
but the cross-over region is much broader than in Fig.~\ref{fig:even}.
}
\label{fig:uneven}
\end{figure}

Physically, $z'(x)$ can be thought of as describing a system 
with overhangs between regions with self-affine scaling,
such as for the spring-network model of Ref.~\cite{GB}, 
or the deposition of a very thin self-affine film onto a stepped surface.
A typical stochastic realization of $z'(x)$ with equally spaced $x_i$ is shown in Fig.~\ref{fig:even}(a),
and the corresponding generalized height-height correlation function, $C_q(r)$,
is shown in Fig.~\ref{fig:even}(b).
Figure~\ref{fig:uneven} shows a similar plot,
but the discontinuity positions, $x_i$, are chosen randomly and uniformly on the interval $(0,1)$.
The number of discontinuities, $N$, is the same for both Fig.~\ref{fig:even} and Fig.~\ref{fig:uneven}.
For length scales $r \gg r_{\times}$, the stepped surface, $z_{\Theta} \equiv z'(x)-z(x)$, 
is expected to be simply a random walk in height with $\alpha_q=0.5$,
but this is not obvious in the numerical data because of the relatively small number of discontinuities in the $x$ interval.

From examination of the numerical results in Figs.~\ref{fig:even} and \ref{fig:uneven}, 
it is obvious that the multi-affinity is caused by the stepped surface, $z_\Theta(x)$,
and the generalized height-height correlation function of $z_\Theta(x)$ can be analytically calculated 
for $r \ll r_{\times}$,
where $r$ is also much smaller than the smallest $x$ separation between discontinuities for finite systems with finite $N$.
\begin{equation}
C_q(r)= \int_0^1 | \sum_{i=1}^N \delta_i \left\{ \Theta(x+r-x_i) - \Theta(x-x_i) \right\} |^q dx \; .
\label{eq:int0}
\end{equation}
The argument, $\Theta(x+r-x_i) - \Theta(x-x_i)$, is either 1 ($x_i - r \le x < x_i$) or 0 (otherwise),
and thus, in the integration range $x=x_i-r$ to $x=x_i$,
only one of the $N$ discontinuities has a non-zero contribution to the integral.
Equation~(\ref{eq:int0}) thus reduces to
\begin{equation}
\begin{array}{lll}
C_q(r)	& = & \sum_{i=1}^N \int_{x_{i-1}}^{x_i} | \sum_{i=1}^N \delta_i \left\{\Theta(x+r-x_i) - \right. \\
	&   & \left. \Theta(x-x_i)\right\}|^q dx \\
	& = & \sum_{i=1}^N \int_{x_i-r}^{x_i} | \delta_i|^q dx \\
	& = & r \sum_{i=1}^N | \delta_i|^q \; .
\end{array}
\label{eq:int1}
\end{equation}

\begin{figure}
\includegraphics[width=0.75\columnwidth]{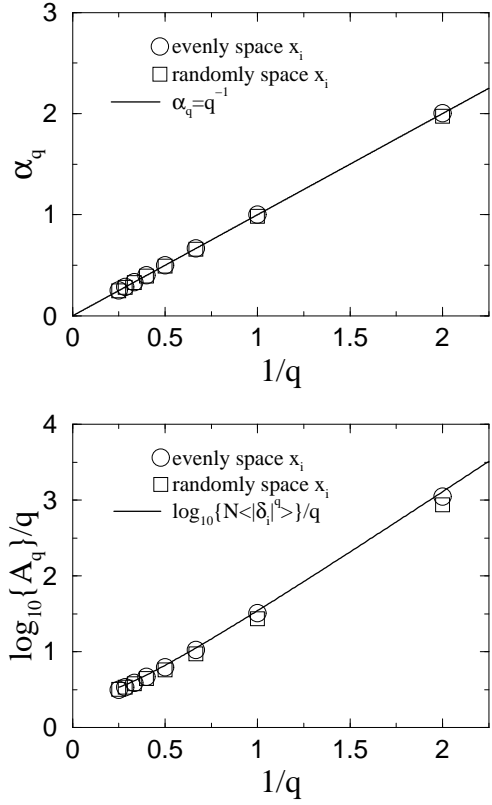}
\caption{Small-$r$ scaling behavior for stepped surfaces, $z_\Theta(x)$, 
shown in Fig.~\ref{fig:even}(b) and Fig.~\ref{fig:uneven}(b).
(a) The multi-affine scaling exponent.
The solid line indicates the analytic solution from Eq.~(\ref{eq:int2}).
(b) The multi-affine scaling prefactor.
The solid line indicates the analytic solution from Eq.~(\ref{eq:int2}).
}
\label{fig:stepped}
\end{figure}

For comparison with Figs.~\ref{fig:even} and \ref{fig:uneven},
\begin{equation}
\frac{1}{q}\log_{10}\{C_q(r)\}  =  \frac{1}{q}\log_{10}\{r\} + \frac{1}{q}\log_{10}\left\{\sum_{i=1}^N | \delta_i|^q\right\} \; ,
\label{eq:int2}
\end{equation}
and thus, $\alpha_q=q^{-1}$ and $\log_{10}\{A_q\} = \log_{10}\left\{N \langle | \delta_i|^q \rangle \right\}$.
See Fig.~\ref{fig:stepped}.
Note that the general form is independent of the $x_i$ distribution
and depends on the $\delta_i$ distribution only through $\langle | \delta_i|^q \rangle$.

\begin{figure}
\includegraphics[width=\columnwidth]{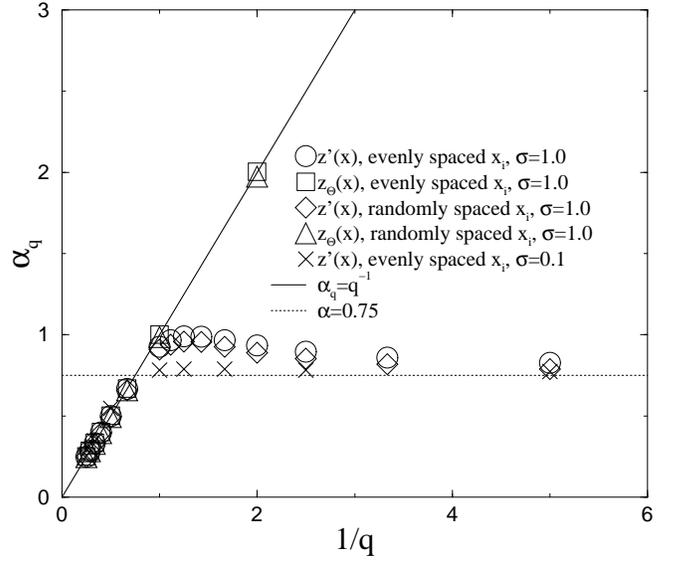}
\caption{Dependence of $\alpha_q$ on $q$ and the relative magnitude of the stepped surface.
The distribution of the discontinuities, $\delta_i$, is Gaussian with mean zero and standard deviation $\sigma$,
and all parameters are the same as in Figs.~\ref{fig:even} and \ref{fig:uneven} unless otherwise indicated.
The lines indicate the two asymptotic behaviors, 
$\alpha_q=1/q$ when $z(x)=0$ and $\alpha_q=0.75$ when $z_{\Theta}(x)=0$.
The data indicated by the symbols are each taken from a single realization of the surface,
and one can see that the spatial distribution of the discontinuities, $x_i$, has no effect on $\alpha_q$.
}
\label{fig:alpha}
\end{figure}

Having derived the small-$r$ scaling behavior for the stepped surface, $z_\Theta(x)$,
it remains to be shown how this multi-affine stepped surface influences the multi-affinity of the complete or mixed surface, $z'(x)$.
Consider the generalized height-height correlation function for the sum of the two surfaces
\begin{equation}
C_q(r)=\langle |\Delta_\Theta^r(x)+\Delta_z^r(x)|^q \rangle \; ,
\label{eq:twofuncs}
\end{equation}
where $\Delta_\Theta^r(x)=z_\Theta(x+r)-z_\Theta(x)$
and $\Delta_z^r(x)=z(x+r)-z(x)$.
The two extremes of $\langle |z(x)| \rangle \ll \langle |z_\Theta(x)| \rangle$ 
and $\langle |z_\Theta(x)| \rangle \ll \langle |z(x)| \rangle$ 
should behave as multi-affine and self-affine surfaces, respectively,
but for intermediate mixed surfaces, the behavior is more complex.
For the numerical results shown in Fig.~\ref{fig:alpha},
two asymptotic scaling regimes are seen.
For large $q$, the mixed surface tends towards multi-affine behavior with $\alpha_q=1/q$,
and for small $q$, the mixed surface tends towards self-affine behavior with $\alpha_q=\alpha$.

We can derive the two asymptotic scaling behaviors for the mixed function by first considering that
for $r \ll r_\times$
\begin{equation}
\begin{array}{lll}
C_q(r)	& = & \langle |\Delta_\Theta^r(x)+\Delta_z^r(x)|^q \rangle \\
	& = & \langle |\Delta_z^r(x) \left(\frac{\Delta_\Theta^r(x)}{\Delta_z^r(x)}+1\right) |^q \rangle \\
	& = & \langle |\Delta_z^r(x)|^q | \frac{\Delta_\Theta^r(x)}{\Delta_z^r(x)}+1 |^q \rangle \\
	& = & \int_0^1 |\Delta_z^r(x)|^q | \frac{\Delta_\Theta^r(x)}{\Delta_z^r(x)}+1 |^q dx\\
	& = & \sum_{i=1}^N \int_{x_{i-1}}^{x_i} |\Delta_z^r(x)|^q | \frac{\Delta_\Theta^r(x)}{\Delta_z^r(x)}+1 |^q dx \; .
\end{array}
\end{equation}
By noticing that $\Delta_\Theta^r(x)=0$ or $\delta_i$ in the interval $x\in (x_{i-1},x_i]$,
\begin{equation}
\begin{array}{lll}
C_q(r)	& = & \sum_{i=1}^N \int_{x_{i-1}}^{x_i-r} |\Delta_z^r(x)|^q | \frac{0}{\Delta_z^r(x)}+1 |^q dx\\
	&   & + \sum_{i=1}^N \int_{x_i-r}^{x_i} |\Delta_z^r(x)|^q | \frac{\delta_i}{\Delta_z^r(x)}+1 |^q dx\\
	& = & \sum_{i=1}^N \int_{x_{i-1}}^{x_i-r} |\Delta_z^r(x)|^q dx\\
	&   & + \sum_{i=1}^N \int_{x_i-r}^{x_i} |\Delta_z^r(x)|^q | \frac{\delta_i}{\Delta_z^r(x)}+1 |^q dx\; .
\end{array}
\end{equation}
As $q \rightarrow 0$, $| \frac{\delta_i}{\Delta_z^r(x)}+1 |^q \approx 1$,
and as $q \rightarrow \infty$, $| \frac{\delta_i}{\Delta_z^r(x)}+1 |^q \approx | \frac{\delta_i}{\Delta_z^r(x)}|^q$.
This gives the following approximations for the two asymptotic regimes,
\begin{equation}
C_q(r) \approx \left\{ 
\begin{array}{ll}
\langle |\Delta_z^r(x)|^q \rangle = A_q r^{q \alpha} & q \ll 1 \\
A_q r^{q \alpha} + r N \langle |\delta_i|^q \rangle & q \gg 1 \; ,
\end{array} \right.
\label{eq:asymp}
\end{equation}
where the additional approximation $\int_{x_{i-1}}^{x_i-r} \cdots dx \approx \int_{x_{i-1}}^{x_i} \cdots dx$ is made,
which is valid when $r \ll 1$.

For large $q$, the scaling may resemble either self-affine scaling or multi-affine scaling,
depending on the exact behavior of $A_q$ for $z(x)$ and the behavior of $\langle |\delta_i|^q \rangle$;
however, for small $q$, the behavior will always resemble self-affine scaling,
provided, of course, that the signal strength of $z(x)$ is sufficiently large compared to the stepped function signal strength
to be numerically noticeable.
For the mixed functions examined in Fig.~\ref{fig:alpha},
self-affine scaling is seen for small $q$
and multi-affine scaling with $\alpha_q=1/q$ is seen for large $q$,
but this is not a universal outcome as indicated in Eq.~(\ref{eq:asymp}).

The scaling behavior of a self-affine surface with vertical discontinuities was investigated numerically and analytically,
and it was shown that the surface of discontinuities (the stepped surface) was the source of the multi-affine behavior.
It was further shown numerically and analytically, 
that the general form for the scaling of the stepped surface at small length scales depends on the distribution of discontinuities
only through $\langle |\delta_i|^q \rangle$.
Two asymptotic scaling behaviors were derived for the self-affine surface with discontinuities,
and for the numerical results shown here, 
self-affine scaling is seen for small $q$
and multi-affine scaling with $\alpha_q=1/q$ is seen for large $q$.
The large-$q$ asymptotic behavior is not universal and depends on the detailed $q$ 
dependence of the mixed function.

These results suggest the need to further study scaling and universality for a variety of systems
where vertical discontinuities are known or are expected to exist.
Such systems include many thin film deposition model and deposition processes onto stepped surfaces.
For these processes, the deposition time should have a large effect on the multi-affine scaling behavior.

\begin{acknowledgments}
The author thanks G.~M.\ Buend\'{\i}a and P.~A.\ Rikvold 
for useful discussions and comments on the manuscript.
Funding was provided by the Netherlands Organization for Scientific Research (NWO).
\end{acknowledgments}


\end{document}